\documentclass[a4paper]{article}

\usepackage{INTERSPEECH2018}

\title{Multimodal speech synthesis architecture for unsupervised speaker adaptation}
\name{Hieu-Thi Luong$^1$ and Junichi Yamagishi$^{1,2}$}
\address{
  $^1$National Institute of Informatics, Tokyo, Japan
  $^2$Unversity of Edinburgh, Edinburgh, UK}
\email{\{luonghieuthi,jyamagis\}@nii.ac.jp}

\begin{document}

\maketitle
\begin{abstract}
This paper proposes a new architecture for speaker adaptation of multi-speaker neural-network speech synthesis systems, in which an unseen speaker's voice can be built using a relatively small amount of speech data without transcriptions. This is sometimes called ``unsupervised speaker adaptation''.
More specifically, we concatenate the layers to the audio inputs when performing unsupervised speaker adaptation while we concatenate them to the text inputs when synthesizing speech from text. Two new training schemes for the new architecture are also proposed in this paper. These training schemes are not limited to speech synthesis; other applications are suggested. Experimental results show that the proposed model not only enables adaptation to unseen speakers using untranscribed speech but it also improves the performance of multi-speaker modeling and speaker adaptation using transcribed audio files.
\end{abstract}

\section{Introduction}

Deep learning is achieving remarkable success in various machine learning tasks and speech synthesis is one such task. In \cite{shen2017natural}, the authors reported that by replacing all traditional processing pipeline including signal processing and text processing with neural networks, the naturalness of synthetic speech can be perceived to be as good as recorded speech. Although this is a significant milestone for the speech synthesis research community, there are still several areas that need to be researched to further advance of text-to-speech (TTS) systems. For example, a state-of-the-art high-quality TTS system is normally built using more than 20 hours of recorded speech data from a professional voice talent and such requirement limits flexible and rapid development of TTS systems due to the high development costs. Studies have shown that statistical parametric speech synthesis systems are capable of modelling the characteristics of multiple voices and adapting models to unseen speaker voices not included in the training database \cite{yamagishi2007average,fan2015multi} by using just a small amount of speech data uttered by each speaker. However, there is still room for improvement and multi-speaker and/or speaker-adaptive neural-network speech synthesis systems are being actively investigated now.

Multi-speaker acoustic models based on neural networks that use augmented speaker identities were also a popular approach for automatic speech recognition \cite{abdel2013fast,miao2014towards,tan2016speaker} as well as speech synthesis \cite{wu2015study,luong2017adapting,doddipatla2017speaker,wan2017integrated,hojo2018dnn,taigman2018voiceloop,ping2018deepvoice3,nachmani2018fitting}. This speaker identity may be represented as a single vector \cite{abdel2013fast,tan2016speaker,wu2015study,luong2017adapting,wan2017integrated,taigman2018voiceloop} or multiple vectors \cite{hojo2018dnn,ping2018deepvoice3}, which could be used as a strategy to increase the number of speaker-specific parameters.
The speaker identity vector could be further categorized into three types: a speaker-code vector (such as one-hot vector, speaker-specific random vector, and speaker-embedded vector \cite{wu2015study,luong2017adapting,hojo2018dnn,ping2018deepvoice3}), an acoustic-driven vector extracted using external models such as i-vector or d-vector \cite{miao2014towards,wu2015study,doddipatla2017speaker}, and an acoustic-driven vector but based on encoders jointly trained with the acoustic model \cite{wan2017integrated,nachmani2018fitting}. 
  
In addition to the modeling of speakers included in the training corpus, some of the architectures can also generate unseen speakers' voices not included in the training corpus. These adaptation schemes share the common goal of fitting a new speaker identity to pre-trained speaker spaces created in the training stage. A new speaker-code vector for the new speaker identity can be estimated on the basic of back-propagation using a small numbers of speech and text pairs of the new speaker \cite{luong2017adapting,2018arXiv180206006A}. In \cite{2018arXiv180206006A}, they attempted to estimate a speaker-code vector from untranscribed speech by training another network separately. An i-vector or d-vector for the new speaker identity can also be computed using external models in the same way as training speakers. The advantage of using the acoustic-driven vector is that we can adapt an acoustic model to an unseen speaker using untranscribed speech \cite{wu2015study,doddipatla2017speaker} with reasonable results. This is sometimes called ``unsupervised speaker adaptation'' \cite{king2008unsupervised,matthew2009unsupervised,DINES2013420}. In these approaches, however, there is an implicit assumption that built neural networks can interpret the speaker space hopefully in the same way as the external models used for extracting the acoustic-driven vector \cite{wu2015study,doddipatla2017speaker}. However, as expected, this assumption does not perfectly hold and the integrated approach usually yields slightly better results. In \cite{wan2017integrated}, an i-vector extractor was jointly optimized with the acoustic model using stochastic gradient descent. In \cite{nachmani2018fitting}, a triplet loss-based speaker encoder was jointly optimized with the acoustic model.

In this paper, motivated by these advancements, we aim to extend our previous speaker adaptation approach based on the speaker-code vector \cite{luong2017adapting} and designing a new architecture that has a similar benefit to the integrated approach. This paper has a similar motivation to \cite{wan2017integrated,nachmani2018fitting,2018arXiv180206006A} but the proposed architecture is different. A key focus for the proposed architecture is to factorize our neural network and learn shared common layers that can be connected to either text or speech modality nets. More specifically, we stack these layers to the speech modality net when we perform speaker adaptation using untranscribed speech to estimate a new speaker code. If adaptation data includes transcriptions, we stack these layers to the text modality net and estimate the speaker code. In both cases, the speaker-code vector is contained within the common layers. In order to synthesize speech from text using a target speaker's voice, we stack the common layers contained the estimated speaker code to the text modality and using them as a conventional acoustic model. Learning neural network layers that can be used for multiple different modal inputs is similar to the concept of \cite{kaiser2017one}. In this paper we use this concept to achieve the speaker adaptation using either transcribed or untranscribed speech from one model, and we further propose a few new training schemes for jointly optimizing the text and speech modality nets. 

The rest of the paper is structured as follows. Section \ref{sec:architecture} describes the proposed architecture. Section \ref{sec:training} outlines a few methodologies to train the proposed model. Experiments and results are presented in Section \ref{sec:experiments}. Section \ref{sec:conclusions} concludes our findings.

\vspace{-1mm}
\section{Multimodal architecture for speaker-adaptive speech synthesis}
\label{sec:architecture}

\begin{figure}[tb]
  \centering
  \includegraphics[width=0.85\columnwidth]{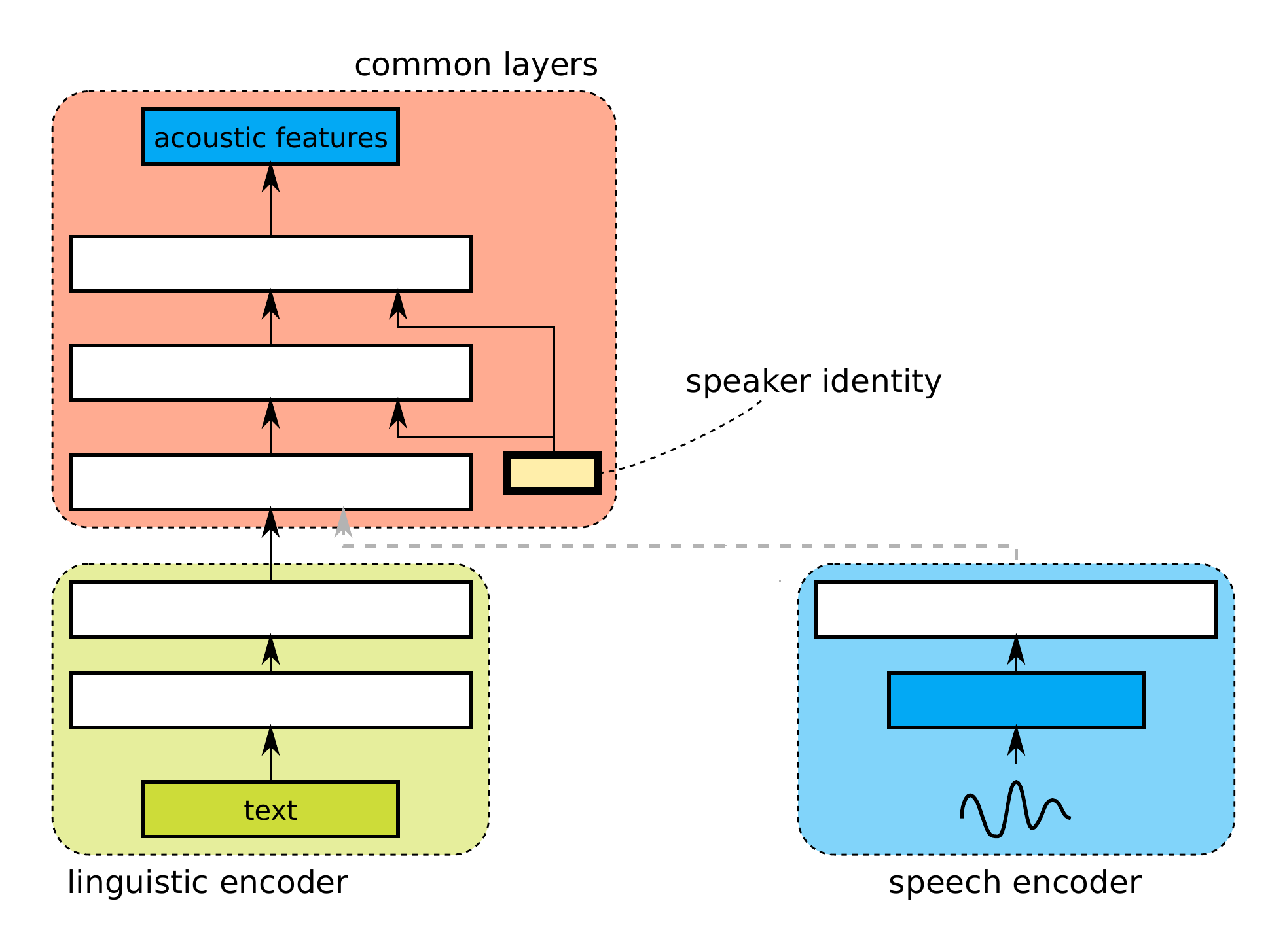}
  \vspace{-3mm}
\caption{Multimodal architecture for multi-speaker acoustic models with auxiliary speech encoder for unsupervised speaker adaptation}
\vspace{-5mm}
\label{fig:architecture}
\end{figure}

Our baseline architecture is a multi-speaker acoustic model using a speaker-embedded vector to present speaker identity \cite{luong2017adapting}. Adaptation to an unseen speaker using transcribed speech is performed by estimating a new embedded vector using back-propagation algorithm to minimize the loss function. This adaptation scheme is essentially a procedure to find the best speaker-embedded vector in a pre-trained speaker space that locally minimizes the distortion between generated and training data. This algorithm is simple but effective \cite{luong2017adapting}. 

To achieve speaker adaptation using untranscribed speech, we factorize the original acoustic model into two components: the common layers and the linguistic encoder as shown in Figure \ref{fig:architecture}. The linguistic encoder can be used together with the common layers to map text to speech. We then introduce a new speech encoder module which could we used in place of the linguistic encoder. The speech encoder can be used with the common layers to map raw waveform data to acoustic features. As the architecture resembles the system investigated in \cite{kaiser2017one}, we call this structure ``multimodal architecture'' and its learning method ``multi-modal learning''\footnote{In the speech processing fields, multi modality means a combination of speech and non-acoustic modality such as images. However, in the proposed model, multi modality indicates different types of input features such as text and signals in the same way as the paper \cite{kaiser2017one}.}. 

In the adaptation stage, if adaptation data includes both speech and its transcription we stack the linguistic encoder and the common layers together. If there are no transcriptions available, we use the speech encoder instead. In either case, the back-propagation algorithm is used to update the speaker-embedded vector for a new speaker. For the former case, we use acoustic-feature and text pairs to compute gradients whereas for the latter case we use acoustic-feature and speech-waveform pairs. 
Note that the speaker identity vector is contained in the common layers rather than in the linguistic or speech encoders.  

\vspace{-1mm}
\section{Multimodal learning methods}
\label{sec:training}

There are several options for training the above multimodal architecture. However, at the time of writing, we believe that these options have not been fully investigated. Therefore we address several strategies for multi-modal learning. We describe two existing naive schemes; and introduce two additional schemes that amend the cost function for the multimodal architecture. The additional schemes work independently and in combination.

\noindent\textbf{Step-by-step training (SS)}: The most common strategy is to first train the text-to-acoustic stack in the usual way, then replace the linguistic encoder with the speech encoder and train the speech encoder while freezing parameters of the common layers. Although this strategy is reasonable, we do not expect it to lead to sufficient results.

\noindent\textbf{Stochastic training}: Similar to an approach used in \cite{kaiser2017one} for training shared components of the multimodal architecture, we could train all the modality layers concurrently by randomly switching input data and replacing modality layers \cite{li2016multi,Wan2017}. Although this strategy would work for our task, due to the large number of factors to be compared in our experiment, we will not investigate this strategy in this paper.

\begin{figure}[tb]
  \centering
  \includegraphics[width=0.63\columnwidth]{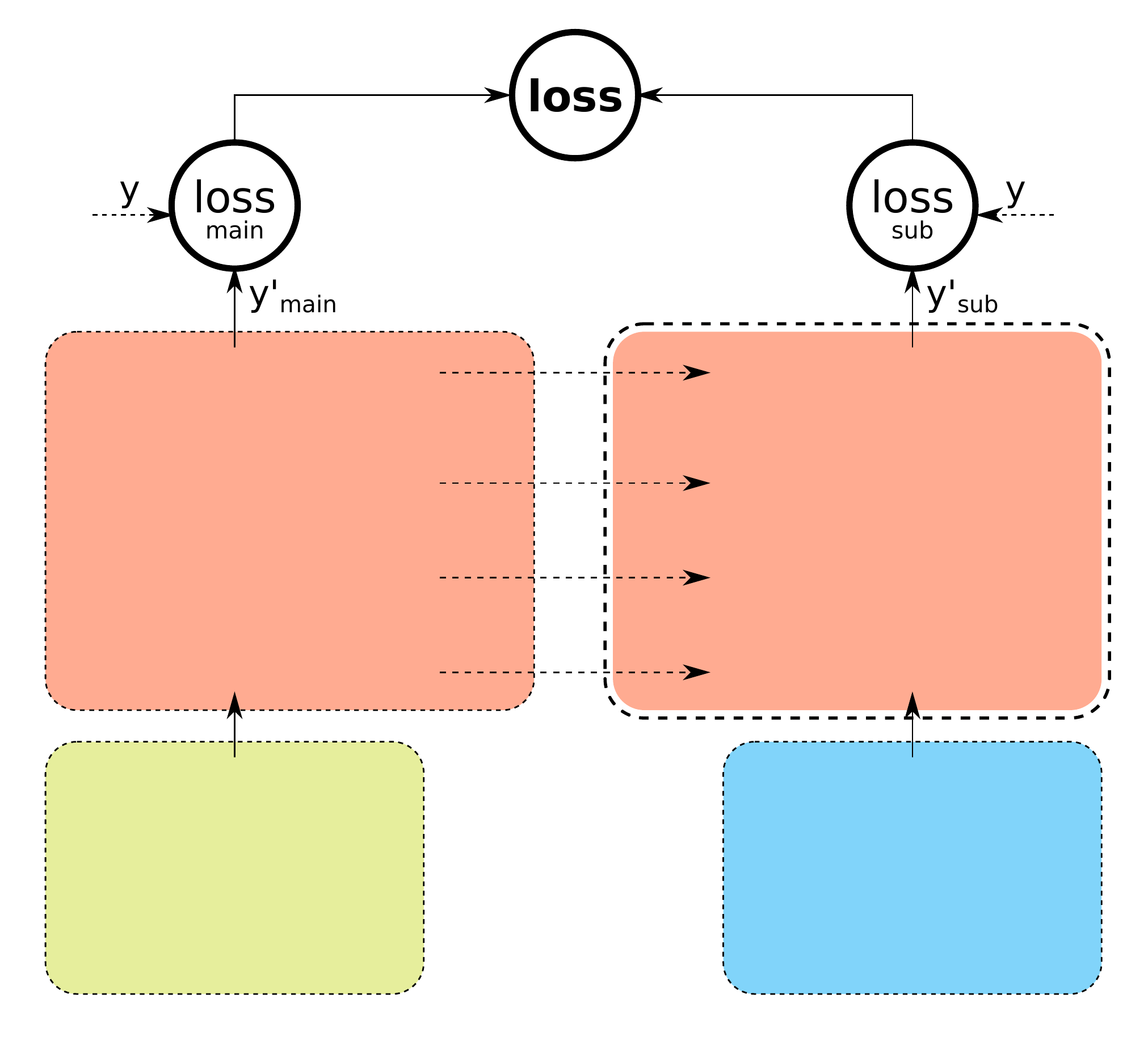} 
  \vspace{-2mm}
\caption{Illustration of the proposed joint-goal training}
\label{fig:training_methods}
\vspace{-5mm}
\end{figure}

\noindent\textbf{Joint goal training with partially shared weights (JG)}: Our first strategy is to learn two networks together. The two networks share the weights of the common layers as illustrated in Figure \ref{fig:training_methods}. One network is attached to the linguistic encoder and the other network is attached to the speech encoder. Unlike the original multi-modal learning in \cite{kaiser2017one} where the goal is to train a single model that could achieve multiple objectives, our architecture prioritizes text-to-speech conversion as the main task and treats speech encoding as the secondary task. We use the following loss function to train the two networks with partially shared weights simultaneously:
\vspace{-1mm}
\begin{equation}
        \textrm{loss} = \textrm{loss}_{main} + \alpha ~ \textrm{loss}_{sub}
\vspace{-1mm}
\end{equation}
where $\alpha$ is a scalar weighting parameter for the loss function of the secondary task. In terms of utilizing the loss function to train multiple objectives, this strategy function similarly to multitask learning \cite{caruana1998multitask}, but, in multitask learning, a mapping is normally one-to-many whereas this mapping is many-to-one. By weighting the sum of the losses while sharing the weights of the common layers, we expect that the models are aware of additional but less important inputs from the secondary network. This acts as a method of regularization to help preserve some weights in the common layers to process the the secondary input, and therefore this would be better than the above stochastic training strategy where we implicitly hope that the common layers can handle both inputs. 
The use of two identical networks together implies a similarity to the Siamese network \cite{bromley1994signature}. However unlike the Siamese network, the motivation of our network is not metric learning. We believe this model can be very useful for other objectives. For instance, by changing the input of the speech encoder network to a different speaker's voice, it is expected that we can train TTS and voice conversion systems that shares the same acoustic feature prediction layers simultaneously. This will be investigated in our future work. 

\begin{figure}[tb]
  \centering
  \includegraphics[width=0.63\columnwidth]{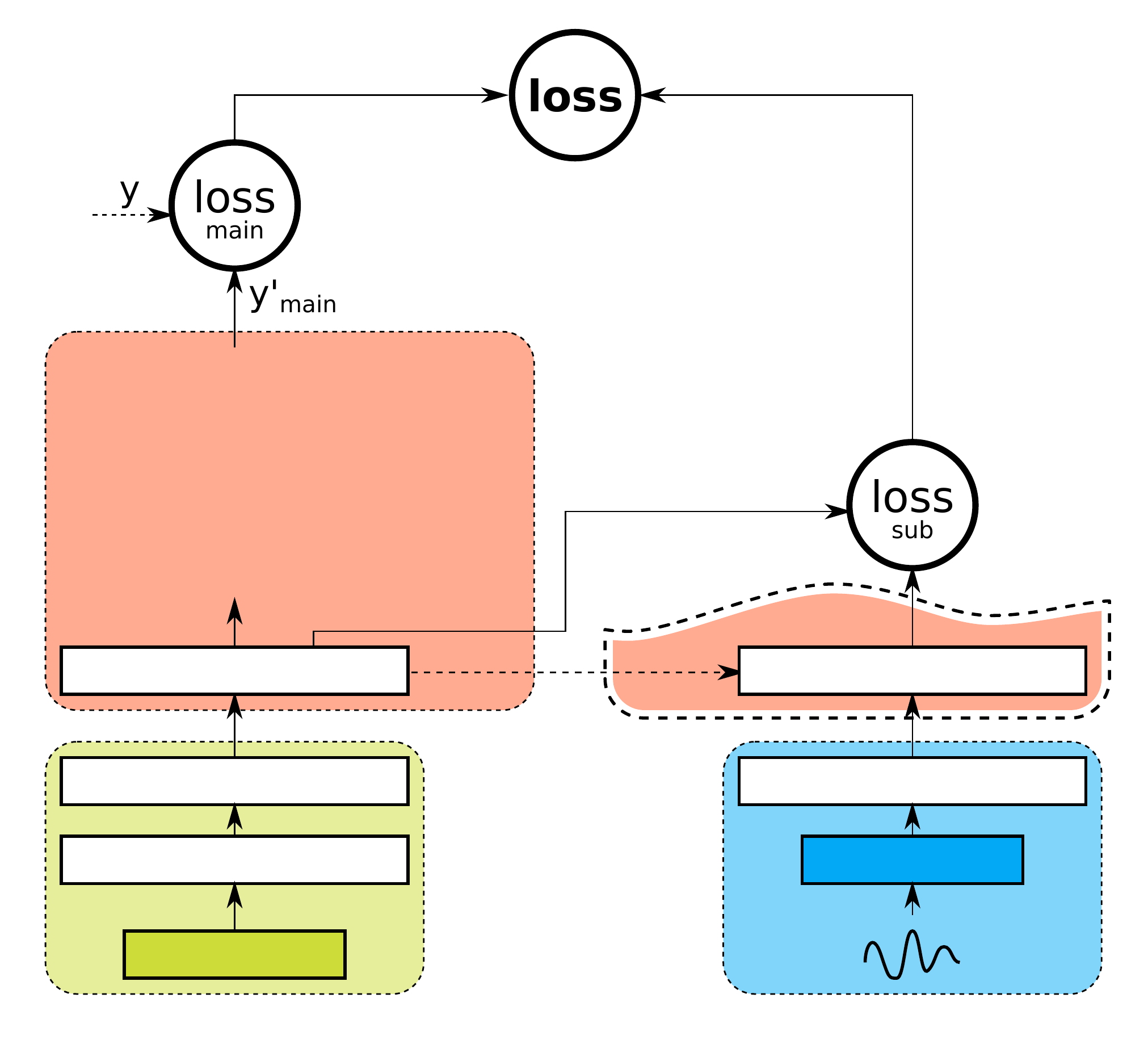} \\ 
  \vspace{-2mm}
\caption{Illustration of the proposed tied-layer training}
\label{fig:training_methods_tl}
\end{figure}

\noindent\textbf{Tied-layers training (TL)}: In the above strategy, we only use the outputs from the two networks. For the next step, we consider to further constrain the inside of the common layers and/or the inputs to the common layers. In particular we want to include outputs of each of
the hidden layers in the common layers stacked on top of text modality net close to its counterpart on top of the speech modality net, as illustrated in Figure \ref{fig:training_methods_tl}. For this purpose, we measure the distance between outputs of the hidden layer in the main network and its counterparts at the secondary network for each training sample and add the distance to the loss function to be minimized:
\vspace{-1mm}
\begin{equation}
    \textrm{loss} = \textrm{loss}_{main} + \beta \sum\limits_l^{L}\textrm{distance}(\boldsymbol{h}^{l}_{main},\boldsymbol{h}^{l}_{sub})
\vspace{-1mm}
\end{equation}
where $\beta$ is a weighting parameter, $L$ is the number of hidden layers to be considered and $\boldsymbol{h}$ represents the outputs of a hidden layer. Example distances would be cosine distance and Euclid distance. We call this strategy ``tied-layers training'' using an analogy from tied-state frameworks for hidden Markov models. 
This idea was inspired by the training scheme of a neural artistic style transfer network \cite{gatys2015neural} in which the models were trained by minimizing loss between hidden layers. 

\noindent\textbf{Combination (JG+TL)}: The above two loss functions complement each other and can be straightforwardly combined in the hope of achieving an even better performance. 
\vspace{-1mm}
\begin{align}
    \textrm{loss} = \textrm{loss}_{main} & + \alpha ~ \textrm{loss}_{sub} \nonumber \\
    & + \beta \sum\limits_l^{L}\textrm{distance}(\boldsymbol{h}^{l}_{main},\boldsymbol{h}^{l}_{sub})
    \vspace{-1mm}
\end{align}

\vspace{-1mm}
\section{Experiments}
\label{sec:experiments}
\subsection{Experimental condition}

\begin{table}[tb]
    \caption{Dataset used for multi-speaker modeling and adaptation experiments.}
    \centering
    \vspace{-2mm}
    \scalebox{0.8}{
    \begin{tabular}{lrrrrrr}
        \hline
        & \multicolumn{3}{c}{Speakers} & \multicolumn{3}{c}{Total utterances} \\ \hline
        & male & female & total & train & valid & test \\ \hline
        Multispeaker & 24 & 20 & 44 & 16,910 & 440 & 440\\
        Adaptation   & 4  & 3  & 7  & vary   & 70  & 70 \\ \hline
    \end{tabular}
    }
    \vspace{-5mm}
    \label{tab:dataset}
\end{table}

We conducted multi-speaker modeling and adaptation experiments using the VCTK corpus \cite{veaux2017cstr}. From the corpus, we chose 51 speakers: 44 were used for training the multi-speaker acoustic models and the rest were used for adaptation. Table \ref{tab:dataset} show the detailed data partitions used for the multi-speaker and adaptation experiments. In the multi-speaker modeling, each speaker had about 400 utterances, while the number of utterances used for adaptation varied from 10 to 320. 
Furthermore, ten utterances were chosen from each speaker for validation and an additional ten utterances were kept as test material.

Standard vocoder features were used as acoustic features in our experiments and the setup is similar to our previous study on a speaker adaptation task \cite{luong2017adapting}. WORLD spectral analysis \cite{morise2016world} was used to obtain smooth spectra from 16 bit speech waveform signals a sampling frequency of 48kHz with a 25 ms window length and 5 ms shift. The obtained features consist of 60-dimensional mel-cepstral coefficients, 25-dimensional band-limited aperiodicities, interpolated logarithm fundamental frequencies and their dynamic features. A voiced/unvoiced binary flag was also included.
English TTS linguistics features were generated using Flite \cite{HTSWorkingGroup2014}. Speech waveform generation was conducted using the WORLD vocoder \cite{morise2016world}. 

Figure \ref{fig:architecture} shows the general setup we used for our experiments.  The text modality net has two feedforward layers while the common layers part has three feedforward layers, followed by a linear output layer to map to the desired dimension. All the feedforward layers contain 1,024 hidden nodes and a sigmoid activation function. For the speech modality net, that is the speech encoder, we used the raw waveform samples as the input. The 16kHz raw waveform was transformed using a 1D convolution layer to the same length as the acoustic output by setting width of convolution to 400 and stride to 80. The convolution layer has 64 filters. The output of the convolution layer was then fed into a 1024-node feedforward layer just before the common layers part. The speaker-embedded vector was set as a 128-dimensional vector; more than twice the number of speakers used in the experiments.

\begin{table}[tb]
    \caption{Models trained to investigate the performance of multi-speaker modeling and speaker adaptation. VL indicates a vanilla TTS system without speech encoder. For TL and JG+TL, only a single common layer at the bottom was tied.}

    \centering
    \vspace{-2mm}
    \scalebox{0.8}{

    \begin{tabular}{clllrr}
        \hline
             & & & & \multicolumn{2}{c}{Loss weight} \\ \hline
         Model  & Strategy       & Speaker-aware & Unsupervised & $\alpha$ & $\beta$  \\ \hline
         VL     & vanilla      & all layers    & no           & -        & - \\
         SS     & step-by-step & last 2 layers & yes          & -        & - \\ 
         JG     & joint goal   & last 2 layers & yes          & 0.5      & - \\
         TL     & tied layers  & last 2 layers & yes          & -        & 1.0 \\
         JG+TL  & combination  & last 2 layers & yes          & 0.2      & 0.2 \\ \hline
    \end{tabular}}
    \vspace{-5mm}
    \label{tab:models}
\end{table}

A total of five models, shown in Table \ref{tab:models}, were trained and evaluated. In the baseline TTS model VL, the speaker-embedded vector was fed into all five feedforward layers. In the remaining models, only the last two feedforward layers of the common layers are speaker-aware layers as illustrated in Figure \ref{fig:architecture}.  For the models using extended loss functions, we used weight values as shown in the table. 
We performed two types of adaptation, supervised adaptation using transcribed speech and unsupervised adaptation using untranscribed speech. The supervised adaptation scheme used was the same back-propagation method as that of \cite{luong2017adapting}. The unsupervised adaptation schemes used the proposed speech encoder, followed by back-propagation to update the speaker-embedded vector. The learning-rate was set to be 0.001 and the Adam optimizer was used. The training was repeated until any loss improvement was not observed over five previous epochs on the validation data or reached a maximum number of epochs (128).

\subsection{Objective evaluation}

\begin{table}[tb]
    \caption{Objective evaluation using mel-cepstral distortion (MCD) in dB and $F_0$ root mean square error ($F_0$ RMSE) for the multi-speaker modeling task. The errors are calculated on the test set including 440 utterances.}
    \centering
    \vspace{-2mm}
    \begin{tabular}{ccr}
        \hline
         Model & MCD [dB] & $F_0$ RMSE  \\ \hline
         VL    & 5.85 & 15.1 \\
         SS    & 5.71 & 15.0 \\ 
         JG    & 5.57 & 14.6 \\
         TL    & 5.79 & 15.3 \\
         JG+TL & 5.63 & 14.8 \\ \hline
    \end{tabular}
    \vspace{-4mm}
    \label{tab:obj_multispeaker}
\end{table}

Table \ref{tab:obj_multispeaker} shows the objective evaluation results of the built models for the multi-speaker task. The table shows Mel-cepstral distortion (MCD) in dB and $F_0$ root mean square error ($F_0$ RMSE). Although the main benefit of the proposed model is unsupervised adaptation, in general, all proposed strategies achieve a similar or even slightly better performance than the baseline system.
%
%
However, since the differences are very small (e.g.\ maximum 0.2 dB in relative in the case of MCD), we conclude that our proposed architecture does not degrade the performance of multi-speaker modeling and provides flexibility to replace the modality nets.

\begin{figure}[tb]
  \centering
  \includegraphics[width=1.0\columnwidth]{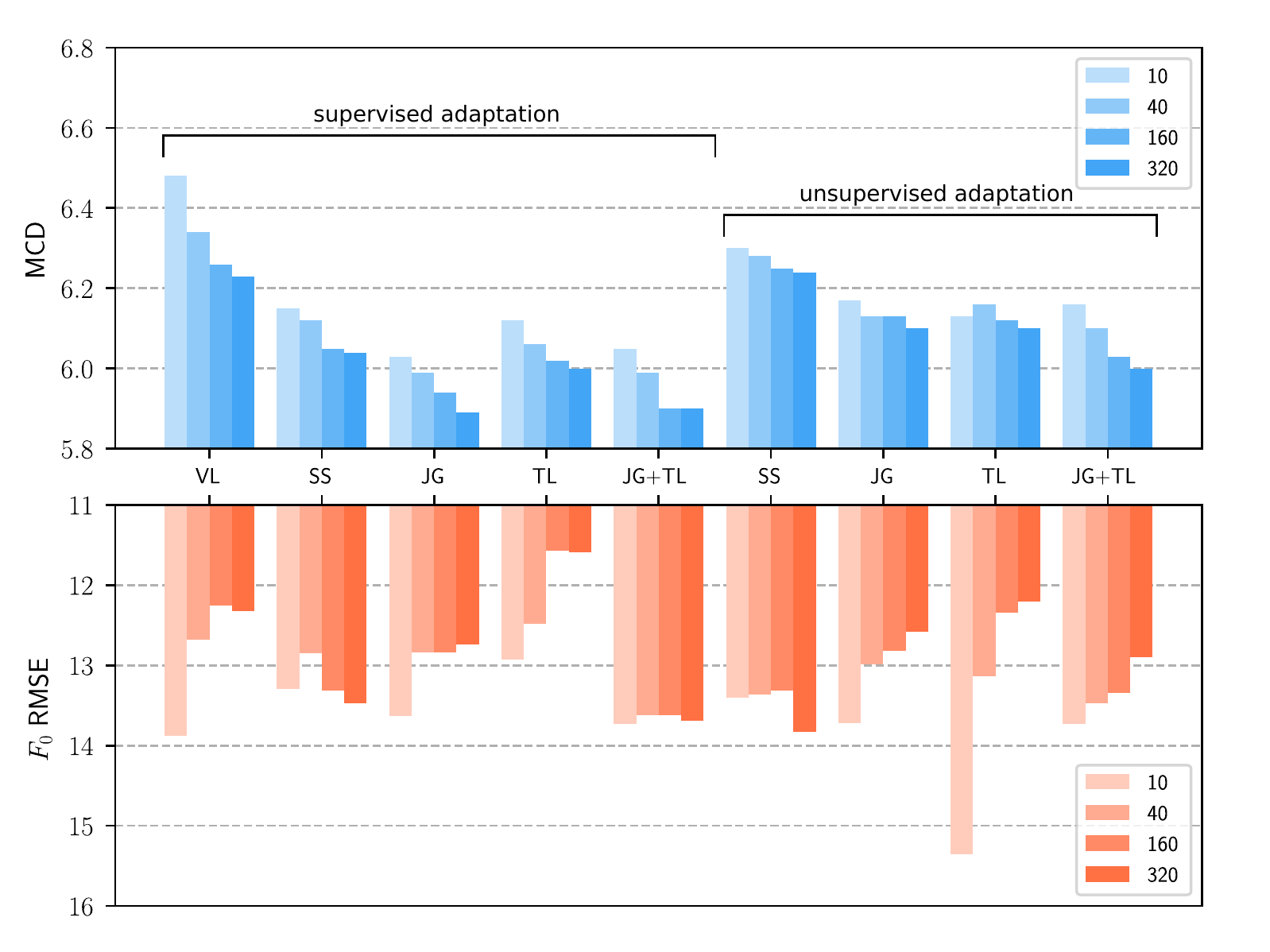}
  \vspace{-5mm}
\caption{Objective evaluation for supervised and unsupervised adaptation tasks.}
\label{fig:obj_adaptation}
\vspace{-5mm}
\end{figure}

Figure \ref{fig:obj_adaptation} shows the objective evaluation results of adaptation using different amounts of data. The results for supervised adaptation and unsupervised adaptation are show on the left and right sides of the figure, respectively.  From the figure we can see that for the supervised adaptation all proposed models yield significantly favorable results over the baseline in terms of MCD. Three models trained using the extended loss functions show further MCD improvements over the SS model. In terms of $F_0$ RMSE, TL shows superior objective results over both the baseline and other proposed models.
The results of unsupervised adaptation  are different to those of their supervised counterpart as expected, but are still positive results. Three models trained using the extended loss functions have smaller errors when the the adaptation data is 40 utterances or above and they are reasonably close to supervised counterparts, which proves the effectiveness of the proposed unsupervised model. 


\subsection{Subjective evaluation}

\begin{figure}[tb]
  \centering
  \includegraphics[width=1.0\columnwidth]{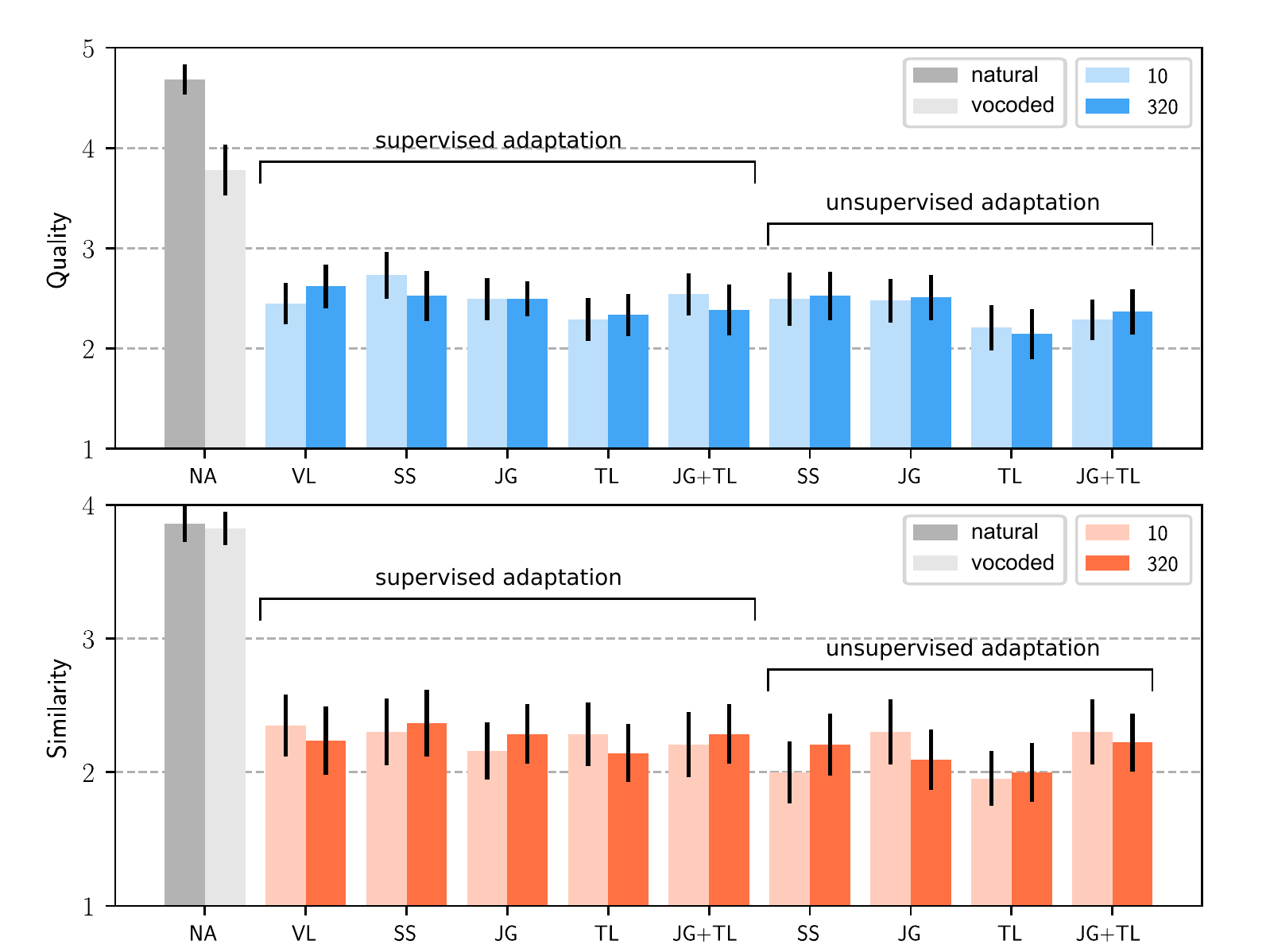}
  \vspace{-5mm}
\caption{Subjective evaluation for supervised and unsupervised adaptation tasks.}
\label{fig:sbj_adaptation}
\vspace{-5mm}
\end{figure}

We conducted a small-scale listening test to subjectively evaluate the proposed model. 29 volunteers participated in two tests consisting of 40 utterances each. The first test used a mean-opinion-score (MOS) on a five-point scale to evaluate the quality of synthetic speech. The second test was a speaker similarity judgment test and the participants were asked to decide if the two samples presented, one synthetic and one natural, of the same sentence are spoken by the same or different speaker and to choose one answer from four options: "Same, absolute sure", "Same, not sure", "Different, not sure" and "Different, absolute sure". The answers were used as a four-point scale measurement.
From the adaptation tasks, we chose four speakers (2 males and 2 females) and evaluated their adapted models using 10 and 320 utterances. We also evaluated their natural and vocoded speech as references. 

The results of the listening test are shown in Figure \ref{fig:sbj_adaptation}. The first system NA indicates the references, that is, recorded speech and vocoded speech. 
The remaining results are either supervised adaptation or unsupervised adaptation using 10 or 320 utterances. 
We observe that the supervised SS system using 10 utterances has the highest MOS score and the unsupervised TL system using 320 utterances has the lowest score. This difference is statistical significant even though the other differences are not. We suspect that the deterministic vocoder did not provide a sufficient quality of speech and our subjects had difficulties to judge small differences of speaker identities. However, we can at least see that unsupervised adaptation has a similar performance to supervised adaptation. We will continue to investigate the proposed model using large-scale datasets and more sophisticated acoustic modeling with the robust speaker-independent Wavenet \cite{juvela2018speaker}.


\section{Conclusions}
\label{sec:conclusions}

In this paper, we proposed a new architecture for speaker adaptation of multi-speaker neural-network speech synthesis systems. They were inspired by the recently proposed multimodal architecture where the modality nets and common body nets are explicitly factorized. For unsupervised adaptation, we proposed to replace a text modality net with a speech modality net so speaker-embedded vectors could be estimated from untranscribed speech data. We also proposed a few training schemes for the multimodal architecture. Experimental results showed that models trained using the proposed loss functions have objective values reasonably close to their supervised counterparts. 
The training schemes proposed in this paper are not limited to speech synthesis. Our future work includes using these schemes for training TTS and voice conversion systems simultaneously.

\noindent\textbf{Acknowledgements:}
This work was partially supported by MEXT KAKENHI Grants (15H01686, 16H06302, 17H04687).

\newpage

\bibliographystyle{IEEEbib}
\bibliography{main}

\begin{thebibliography}{10}

\bibitem{shen2017natural}
Jonathan Shen, Ruoming Pang, Ron~J. Weiss, Mike Schuster, Navdeep Jaitly,
  Zongheng Yang, Zhifeng Chen, Yu~Zhang, Yuxuan Wang, RJ~Skerry-Ryan, Rif~A.
  Saurous, Yannis Agiomyrgiannakis, and Yonghui Wu,
\newblock ``Natural {TTS} synthesis by conditioning {WaveNet} on {M}el
  spectrogram predictions,''
\newblock in {\em Proc. ICASSP}, 2018.

\bibitem{yamagishi2007average}
Junichi Yamagishi and Takao Kobayashi,
\newblock ``Average-voice-based speech synthesis using {HSMM}-based speaker
  adaptation and adaptive training,''
\newblock {\em IEICE T.\ Inf.\ Syst.}, vol. 90, no. 2, pp. 533--543, 2007.

\bibitem{fan2015multi}
Yuchen Fan, Yao Qian, Frank~K. Soong, and Lei He,
\newblock ``Multi-speaker modeling and speaker adaptation for {DNN}-based {TTS}
  synthesis,''
\newblock in {\em Proc. ICASSP}, 2015, pp. 4475--4479.

\bibitem{abdel2013fast}
Ossama Abdel-Hamid and Hui Jiang,
\newblock ``Fast speaker adaptation of hybrid {NN}/{HMM} model for speech
  recognition based on discriminative learning of speaker code,''
\newblock in {\em Proc. ICASSP}, 2013, pp. 7942--7946.

\bibitem{miao2014towards}
Yajie Miao, Hao Zhang, and Florian Metze,
\newblock ``Towards speaker adaptive training of deep neural network acoustic
  models,''
\newblock in {\em Proc. Interspeech}, 2014.

\bibitem{tan2016speaker}
Tian Tan, Yanmin Qian, Dong Yu, Souvik Kundu, Liang Lu, Khe~Chai Sim, Xiong
  Xiao, and Yu~Zhang,
\newblock ``Speaker-aware training of {LSTM}-{RNN}s for acoustic modelling,''
\newblock in {\em Proc. ICASSP}, 2016, pp. 5280--5284.

\bibitem{wu2015study}
Zhizheng Wu, Pawel Swietojanski, Christophe Veaux, Steve Renals, and Simon
  King,
\newblock ``A study of speaker adaptation for {DNN}-based speech synthesis,''
\newblock in {\em Proc. Interspeech}, 2015, pp. 879--883.

\bibitem{luong2017adapting}
Hieu-Thi Luong, Shinji Takaki, Gustav~Eje Henter, and Junichi Yamagishi,
\newblock ``Adapting and controlling {DNN}-based speech synthesis using input
  codes,''
\newblock in {\em Proc. ICASSP}, 2017, pp. 4905--4909.

\bibitem{doddipatla2017speaker}
Rama Doddipatla, Norbert Braunschweiler, and Ranniery Maia,
\newblock ``Speaker adaptation in {DNN}-based speech synthesis using
  d-vectors,''
\newblock in {\em Proc. Interspeech}, 2017, pp. 3404--3408.

\bibitem{wan2017integrated}
Moquan Wan, Gilles Degottex, and Mark~J.F. Gales,
\newblock ``Integrated speaker-adaptive speech synthesis,''
\newblock in {\em Proc. ASRU}, 2017, pp. 705--711.

\bibitem{hojo2018dnn}
Nobukatsu Hojo, Yusuke Ijima, and Hideyuki Mizuno,
\newblock ``{DNN}-based speech synthesis using speaker codes,''
\newblock {\em IEICE T. Inf. Syst.}, vol. 101, no. 2, pp. 462--472, 2018.

\bibitem{taigman2018voiceloop}
Yaniv Taigman, Lior Wolf, Adam Polyak, and Eliya Nachmani,
\newblock ``Voiceloop: Voice fitting and synthesis via a phonological loop,''
\newblock in {\em Proc. ICLR}, 2018.

\bibitem{ping2018deepvoice3}
Wei Ping, Kainan Peng, Andrew Gibiansky, Sercan~O Arik, Ajay Kannan, Sharan
  Narang, Jonathan Raiman, and John Miller,
\newblock ``Deep voice 3: 2000-speaker neural text-to-speech,''
\newblock in {\em Proc. ICLR}, 2018.

\bibitem{nachmani2018fitting}
Eliya Nachmani, Adam Polyak, Yaniv Taigman, and Lior Wolf,
\newblock ``Fitting new speakers based on a short untranscribed sample,''
\newblock {\em arXiv preprint arXiv:1802.06984}, 2018.

\bibitem{2018arXiv180206006A}
Sercan~O. Arik, Jitong Chen, Peng Kainan, Wei Ping, and Yanqi Zhou,
\newblock ``{Neural Voice Cloning with a Few Samples},''
\newblock {\em ArXiv e-prints}, Feb. 2018.

\bibitem{king2008unsupervised}
Simon King, Keiichi Tokuda, Heiga Zen, and Junichi Yamagishi,
\newblock ``Unsupervised adaptation for {HMM}-based speech synthesis,''
\newblock in {\em Proc. Interspeech}, 2008.

\bibitem{matthew2009unsupervised}
Matthew Gibson,
\newblock ``Two-pass decision tree construction for unsupervised adaptation of
  {HMM}-based synthesis models,''
\newblock in {\em Proc. Interspeech}, 2009.

\bibitem{DINES2013420}
John Dines, Hui Liang, Lakshmi Saheer, Matthew Gibson, William Byrne, Keiichiro
  Oura, Keiichi Tokuda, Junichi Yamagishi, Simon King, Mirjam Wester, Teemu
  Hirsimäki, Reima Karhila, and Mikko Kurimo,
\newblock ``Personalising speech-to-speech translation: Unsupervised
  cross-lingual speaker adaptation for hmm-based speech synthesis,''
\newblock {\em Computer Speech \& Language}, vol. 27, no. 2, pp. 420 -- 437,
  2013.

\bibitem{kaiser2017one}
Lukasz Kaiser, Aidan~N Gomez, Noam Shazeer, Ashish Vaswani, Niki Parmar, Llion
  Jones, and Jakob Uszkoreit,
\newblock ``One model to learn them all,''
\newblock {\em arXiv preprint arXiv:1706.05137}, 2017.

\bibitem{li2016multi}
Bo~Li and Heiga Zen,
\newblock ``Multi-language multi-speaker acoustic modeling for {LSTM-RNN} based
  statistical parametric speech synthesis.,''
\newblock in {\em Proc. Interspeech}, 2016, pp. 2468--2472.

\bibitem{Wan2017}
Vincent Wan, Yannis Agiomyrgiannakis, Hanna Silen, and Jakub Vit,
\newblock ``Google’s next-generation real-time unit-selection synthesizer
  using sequence-to-sequence {LSTM}-based autoencoders,''
\newblock in {\em Proc.\ Interspeech}, 2017, pp. 1143--1147.

\bibitem{caruana1998multitask}
Rich Caruana,
\newblock ``Multitask learning,''
\newblock in {\em Learning to learn}, pp. 95--133. Springer, 1998.

\bibitem{bromley1994signature}
Jane Bromley, Isabelle Guyon, Yann LeCun, Eduard S{\"a}ckinger, and Roopak
  Shah,
\newblock ``Signature verification using a "{Siamese}" time delay neural
  network,''
\newblock in {\em Advances in Neural Information Processing Systems}, 1994, pp.
  737--744.

\bibitem{gatys2015neural}
Leon~A Gatys, Alexander~S Ecker, and Matthias Bethge,
\newblock ``A neural algorithm of artistic style,''
\newblock {\em arXiv preprint arXiv:1508.06576}, 2015.

\bibitem{veaux2017cstr}
Christophe Veaux, Junichi Yamagishi, Kirsten MacDonald, et~al.,
\newblock ``{CSTR} {VCTK} corpus: English multi-speaker corpus for {CSTR}
  {V}oice {C}loning {T}oolkit,'' 2017.

\bibitem{morise2016world}
Masanori Morise, Fumiya Yokomori, and Kenji Ozawa,
\newblock ``{WORLD}: a vocoder-based high-quality speech synthesis system for
  real-time applications,''
\newblock {\em IEICE T.\ Inf.\ Syst.}, vol. 99, no. 7, pp. 1877--1884, 2016.

\bibitem{HTSWorkingGroup2014}
{HTS Working Group},
\newblock ``The {English TTS} system {Flite}+{HTS}\_engine,'' 2014.

\bibitem{juvela2018speaker}
Lauri Juvela, Vassilis Tsiaras, Bajibabu Bollepalli, Manu Airaksinen, Junichi
  Yamagishi, and Paavo Alku,
\newblock ``Speaker-independent raw waveform model for glottal excitation,''
\newblock in {\em Proc.\ Interspeech}, 2018.

\end{thebibliography}
\end{document}